# Strong magneto-elastic coupling and polar properties in orthorhombic $Eu_{1-x}Y_xMnO_3$ manganite

J. AGOSTINHO MOREIRA, A. ALMEIDA, W. S. FERREIRA, J. P. ARAÚJO, A. M. PEREIRA AND M. R. CHAVES

*IFIMUP and IN- Institute of Nanoscience and Nanotechnology. Departamento de Física e Astronomia da Faculdade de Ciências da Universidade do Porto. Rua do Campo Alegre, 687. 4169-007 Porto. Portugal.*

M. M. R. COSTA AND V. A. KHOMCHENKO

*Centro de Estudos de Materiais por Difracção de Raios-X. Departamento de Física da Faculdade de Ciências e Tecnologia da Universidade de Coimbra, Rua Larga, 3004-516 Coimbra, Portugal*

J. KREISEL

*Laboratoire des Matériaux et du Génie Physique, Minatec, Grenoble Institute of Technology, CNRS, 38016 Grenoble, France.*

D. CHERNYSHOV

*Swiss-Norwegian Beam Lines at the European Synchrotron Radiation Facility (ESRF), 38000, Grenoble, France*

S. M. F. VILELA AND P. B. TAVARES

*Centro de Química - Vila Real. Universidade de Trás-os-Montes e Alto Douro. Apartado 1013, 5001-801. Vila Real. Portugal.*

*e-mail: jamoreir@fc.up.pt

## ABSTRACT

This work reports an experimental investigation of the magneto-elastic coupling and polarization character of magnetic phases of the orthorhombic $Eu_{1-x}Y_xMnO_3$ system at low temperatures.

The temperature dependence of the polarization reversal curves clearly reveals the existence of a reentrant improper ferroelectric phase for $x$ = 0.2 and 0.3. Although a ferroelectric phase is also stable for $x$ = 0.4, we have no experimental evidence that it vanishing at finite temperatures. From these results and those obtained from other experimental techniques, the corresponding ($x$,T) phase diagram was traced, yet yielding significant differences with regard to previous reports.

An expressive magneto-elastic coupling is revealed by changes observed in both Mn-O bond lengths and Mn-O1-Mn bond angle at the magnetic phase transitions, obtained by temperature dependence of synchrotron x-ray diffraction experiments. Furthermore, signatures of the lattice deformations across the magnetic phase transitions were evidenced by anomalies in the temperature dependence of the lattice mode involving rotations of the $MnO_6$ octahedra. These anomalies confirm the important role of the spin-phonon coupling in these materials.

# I. INTRODUCTION

Magnetoelectric compounds have attracted a lot of interest in the scientific community due to the coupling between electric polarization and magnetic order.[1-4] Particularly, multiferroic materials, which exhibit ferromagnetism and ferroelectricity coupled together in the same thermodynamic phase, are of considerable interest due to their potential applicability in novel technological devices, opening the possibility to control spin transport electrically.[5,6] In earlier published works, it has been evidenced that the mechanisms underlying magnetoelectricity and multiferroicity are complex, involving both competitive magnetic interactions and spin-phonon coupling.[7-10] Despite the intensive experimental studies performed in several families of magnetoelectric compounds, no universal model was proposed. In fact, it has been assumed that the inverse Dzyaloshinskii-Moriya interaction induces the electric polarization of the electronic orbitals, without the involvement of the lattice degrees of freedom.[11] An alternative model was proposed in Ref. 12, which states that the Dzyaloshinskii-Moriya interaction has two different effects: it induces the ferroelectric state through lattice deformations, and stabilizes the magnetic structure at low temperatures.

Among rare-earth manganites, orthorhombic $GdMnO_3$, $TbMnO_3$ and $DyMnO_3$ have shown magnetoelectric properties.[14] The sinusoidal antiferromagnetic order in $TbMnO_3$ and $DyMnO_3$, is considered to be responsible for the modulation of the Mn-O1-Mn angle, which has been taken as primary order parameter in these improper ferroelectrics.[7] Goto *et al*[7] have stressed the importance of the Mn-O1-Mn bond angle conditioning the orbital overlapping and, thus, underlying the physical properties of the phase sequences at low-temperatures. It has been found that the superexchange integrals of nearest neighbor (NN) ferromagnetic ($J_1 < 0$) and next-nearest neighbor (NNN) antiferromagnetic ($J_2 > 0$) interactions between the Mn spins depend strongly on the Mn-O1-Mn angle.[7] Namely, it has been shown that with decreasing the Mn-O1-Mn bond angle, $|J_1|$ decreases and $|J_2|$ increases, due to a large overlap integral between the two oxygen 2p orbitals along the a + b axis.[7] Therefore, the system is regarded as a frustrated spin system described by the so-called $J_1$-$J_2$ localized spin model. The systematic change of magnetic ordering induced by a change of the rare-earth ion in $ReMnO_3$ is understood by the frustrated model; the A-type antiferromagnetic order arises mainly from the NNN antiferromagnetic interaction $J_2$. In the magnetically-driven ferroelectrics, lattice distortion, owing to the coupling with magnetic momenta, has been considered to be the very origin of the spontaneous polarization apparent in the temperature range of stability of the ferroelectric phase. Attempts have been undertaken to establish the order of magnitude of those distortions, but one major handicap has been their small size, which is expected to be three or more orders of magnitude lower than that of typical ferroelectrics like $BaTiO_3$.[15-18] Surprisingly though, recently published results for the hexagonal $YMnO_3$ and $LuMnO_3$ showed unusual large displacements near $T_N$, giving rise to a giant magneto-elastic coupling in these materials.[19] Yet, the ferroelectric polarization remains almost constant through $T_N$. Unlike the usual ferroelectrics, the isostructural transition of the referred hexagonal compounds at $T_N$ is not accompanied by a soft-mode.

A detailed and systematic experimental study of the mechanisms associated with the coupling between magnetic order and electric polarization in rare-earth manganites is still missing. The aforementioned $GdMnO_3$, $TbMnO_3$, and $DyMnO_3$ compounds are not the best candidates for such a study, because changes on the Mn-O1-Mn angle are also affected by changes on the magnetic momenta of the unit cell whenever a rare-earth ion is replaced by another. On the other hand, the possibility of systematic and fine tuning of the A-site size, without increasing the magnetic complexity arising from the rare-earth ion, can be achieved by the isovalent substitution of the trivalent $Eu^{3+}$ ion by $Y^{3+}$ in $Eu_{1-x}Y_xMnO_3$, for $x < 0.55$. This solid solution allows expecting a continuous variation of the Mn-O1-Mn bond angle, which is associated with the development of the complex magnetic ground states and ferroelectric phases, analogous to the non-doped $GdMnO_3$, $TbMnO_3$ and $DyMnO_3$.

Earlier literature reports on $Eu_{1-x}Y_xMnO_3$, with $0 \leq x < 0.55$, have already shown that this system exhibits a complex and interesting phase diagram with various magnetic and ferroelectric phases, making it an attractive system to study coupling between polar, magnetic and structural degrees of freedom.[20-24] Notwithstanding numerous studies, the phase diagram is still not unambiguously established (see section II for more details). Moreover, despite the observation of spin-phonon coupling and the existence of electromagnons in $Eu_{1-x}Y_xMnO_3$, evidenced through both Raman scattering and infrared spectroscopies, [21,25-28] no information regarding the lattice deformations occurring through the low temperature magnetic phase transitions, as well as their contribution to the spin-lattice coupling have yet been reported for this system. The establishment of an electric polarization implies a loss of the inversion symmetry centre and lattice deformations, arising from atomic polar displacements. As a consequence, a carefully study of the crystallographic structure across the phase transitions is desirable.

In this paper we present a detailed study of the crystal structure and polar properties of orthorhombic $Eu_{1-x}Y_xMnO_3$, in the concentration range $0 \leq x \leq 0.4$, by using both synchrotron radiation powder diffraction and polarization versus electric field (P(E)) relations. We aim at correlating the temperature dependence of lattice distortions with the magnetic phase transition sequence, and, in this way, getting further information regarding the coupling between the spin system and the lattice, which yields large magneto-elastic coupling in these compounds.

## II. General considerations and the phase diagram of $Eu_{1-x}Y_xMnO_3$, $x < 0.55$

In this section, we critically review the literature on the phase diagram of $Eu_{1-x}Y_xMnO_3$, $x < 0.55$, as it will be crucial for the understanding of our main outcomes of this work.

$Eu_{1-x}Y_xMnO_3$, with $0 \leq x < 0.55$, exhibits distinctive features making it an attractive system to study. In this system, physical properties are driven by the magnetic spin of the $Mn^{3+}$ ions, but they can be drastically changed by varying the content of $Y^{3+}$, which does not carry any magnetic moment but changes the effective A-site size, and, thus, the Mn-O1-Mn bond angle.[21]

The phase diagram of $Eu_{1-x}Y_xMnO_3$, with $0 \leq x < 0.55$, has been described on the grounds of competitive NN ferromagnetic and NNN antiferromagnetic interactions, along with single-ion anisotropy and the Dzyaloshinskii-Moriya interaction.[29] As a consequence, these compounds exhibit a rich variety of phase transitions from incommensurate to commensurate antiferromagnetic phases, some of them with a ferroelectric character, depending on the extent of chemical substitution $x$.

Ivanov *et al*[22], Hemberger *et al*[23], and Yamasaki *et al*[24] have proposed ($x$,T) phase diagrams, for $Eu_{1-x}Y_xMnO_3$ single crystals, with $0 \leq x < 0.55$, obtained by using both identical and complementary experimental techniques. Although the proposed phase diagrams present discrepancies regarding the magnetic phase sequence and the ferroelectric properties for $0.15 < x < 0.25$, there is a good agreement concerning the phase sequence for $0.25 < x < 0.55$.

For all compounds, the paramagnetic-paraelectric state above $T_N \approx 50 - 45$ K, is followed by an ordered antiferromagnetic phase (AFM-1) with incommensurate modulation of the manganese spins. According to Hemberger *et al*,[23] for $x < 0.15$, a weakly ferromagnetic phase is established, with a canted A-type antiferromagnetic order (AFM-3). On the other hand, Yamasaki *et al*[24] have reported the spread of this phase over $x$ up to 0.25. Due to the collinear structure, this antiferromagnetic phase is not ferroelectric in the absence of an applied magnetic field. Very recently, Tokunaga *et al*[30,31] have published a high magnetic field study of the polar properties of $EuMnO_3$ and $Eu_{0.9}Y_{0.1}MnO_3$. Application of magnetic fields parallel to the *b*-axis, as high as 20 T, causes first order transition to the ferroelectric phase, with polarization parallel to the *a*-axis, below 44 K.

For higher concentrations $x > 0.25$, the ground state is antiferromagnetic (AFM-2) and ferroelectric, without a ferromagnetic component.[23,24] The modulation vector (0 $q_l$ 0) persists down to 5 K, and the magnetic structure is cycloidal.[24] For x > 0.35, two successive ferroelectric phase transitions occur, with spontaneous polarization along the *c*-axis and *a*-axis, respectively.[23,24] In agreement with the inverse Dzyaloshinskii-Moriya model, the direction of the electric polarization is associated with the *bc*- and *ab*-cycloidal states, which can be reversed by an applied magnetic field.[32] In these ferroelectric phases, a remarkable magnetic anisotropy is found. A microscopic model has shown that the Dzyaloshinskii-Moriya interaction between spins neighboring along the cubic x and y axes, as well as, the single-ion anisotropy favor the *ab*-cycloidal spin state with P//a, while the Dzyaloshinskii-Moriya interactions between spins neighboring along the c-axis favor the *bc*-cycloidal state with P//c. Their competition is controlled by the NNN $J_2$ exchanges enhanced by lattice distortion, This leads to a lattice-distortion-induced reorientation of polarization from *c* to *a* with decreasing the A-site radius.[29]

The phase diagram of the range of compositions $0.15 < x < 0.25$ is still controversial. For the special case of $x = 0.20$, Hemberger *et al*[23] detected in both specific heat and electric permittivity hints for another phase transition at $T_{AFM-2}$ = 30 K. Double magnetic hysteresis loops at 25 K were reported, revealing the antiferromagnetic character of the phase below $T_{AFM-2}$.[23] Based on the anomalous behaviour observed in the electric permittivity and magnetization curves, a canted antiferromagnetic phase (AFM-3) below $T_{AFM-3}$ = 22 K was proposed.[23] According to Hemberger *et al*,[23] $Eu_{0.8}Y_{0.2}MnO_3$ becomes ferroelectric below $T_{AFM-2}$ = 30 K along the *a*-direction. The ferroelectric character of both low temperature AFM-2 and AFM-3 magnetic phases was also found by Valdés *et al* in $Eu_{0.75}Y_{0.25}MnO_3$.[26] Taking into account the weak ferromagnetic character of the AFM-3 phase, as well as the electric polarization below $T_{AFM-2}$ = 30K, Hemberger *et al*[23] have proposed a non-collinear spiral order for the AFM-2 phase, and a spin-canting cone-like structure, for the AFM-3 one. The actual magnetic ordering of the low temperature phases, however, remains still unknown. On the other hand, Yamasaki *et al*[24] have reported another phase sequence for $Eu_{0.8}Y_{0.2}MnO_3$ from the incommensurate magnetic phase to the canted A-type antiferromagnetic one (AFM-2), without any ferroelectric properties. Very recently, a carefully study of the P(E) hysteresis loops and pyroelectric current has shown that the ferroelectric character of the $Eu_{0.8}Y_{0.2}MnO_3$ only appears between $T_{AFM-2}$ and $T_{AFM-3}$; i.e., in the AFM-2 phase, with very small values of the spontaneous polarization, arising from lattice deformations underlined by the microscopic mechanisms associated with the phase transition at $T_{AFM-2}$.[21,33]

The origin of the ferroelectricity in these compounds could be understood in the framework of the spin-driven ferroelectricity model. In these frustrated spin systems, the inverse Dzyaloshinskii-Moriya interaction mechanism has been proposed.[11] However, based on experimental results, the magnetic structure has been well established only for the compositions $x = 0.4$ and 0.5. For the other compositions, the magnetic structure is not yet elucidated.

The ferroelectric properties of the $Eu_{1-x}Y_xMnO_3$ have also been studied by measurement of the electric current after cooling the sample under rather high-applied electric fields (E > 1 kV/cm). As it was shown in Refs. 33 and 35, $Eu_{1-x}Y_xMnO_3$ exhibits a rather high polarizability, which can hinder the actual spontaneous polarization to be ascertained. In this case, a special experimental procedure has to be carried out.[33,35]

## III. EXPERIMENTAL

High quality $Eu_{1-x}Y_xMnO_3$ ceramics were prepared by the sol-gel urea combustion method. A detailed study of $EuMnO_3$ and $GdMnO_3$ ceramics prepared in this way, has lead to results very similar to the ones obtained in the corresponding single crystals.[20]

The valence of the europium ion was checked through x-ray photoemission spectroscopy (XPS) technique, and no evidences of the existence of valences other than the Eu (III) could be detected. As the samples were fast cooled from 1350 ºC down to room temperature, significant deviations of the oxygen occupancy from the expected stoichiometric $Eu_{0.8}Y_{0.2}MnO_3$ are not expected, thus excluding the existence of significant amount of Mn (IV) ion.[36] The Rietveld refinement of powder x-ray diffraction data revealed no impurity phases. Refined occupancies of crystallographic positions were found to correspond to the nominal compositions of the samples.

The dielectric and magnetic properties of all samples prepared were previously studied and compared with the published data.[20,34,35] No significant differences between the temperature dependence of the dielectric constant and of the specific induced magnetization obtained in this work and those published in current literature could be detected. These results, along with the data obtained from the x-ray diffraction and energy dispersion spectroscopy attests the high quality of our samples.

Rectangular shaped samples were prepared from the ceramic pellet, and gold electrodes were deposited using the sputtering method. The P(E) relation was recorded between 50 K and 8 K, using a modified Sawyer-Tower circuit.[37] In order to prevent any dynamical response, from masking the actual domain reversal, we have chosen to perform the measurements of the P(E) at enough low operating frequencies. As the P(E) relations do not change with frequency below 1 Hz, we have taken 330 mHz as the operating frequency.

Synchrotron x-ray diffraction experiments were performed at the beamline BM01A (Swiss-Norwegian BeamLines) at the European Synchrotron Radiation Facility (ESRF), Grenoble, France. For each composition, a powder sample prepared with uniform grain size was enclosed in a sealed Lindemann glass capillary (0.2mm diameter to minimize the absorption factor, $\rho.\mu.r$ and 0.01mm wall thickness) and mounted on an appropriate golden plate sample holder to be attached to the cryostat.

Diffraction data were collected using a wavelength of $\lambda$=0.70Å from a Si(111) double crystal monochromator and a MAR345 image plate detector at a sample-to-detector distance of 220 mm. This setup was found to be a good compromise between resolution and accessible range of *d*-spacing. Calibration was performed by using a $LaB_6$ powder as a reference material (NIST standard reference material 660a). The powder sample remained stationary during data collection. The exposure times for each data set were selected between 10 and 20 seconds to avoid oversaturation of the detector. Measurements were made at temperatures between 5 K and 290 K. The samples were cooled in a flow He cryostat.

Preliminary data reduction, including reconstruction of 2D diffraction patterns from the raw data, was carried out using the ESRF Fit2D software,[38] yielding intensity versus $2\theta$ diffraction pattern. Calculation of standard deviations and scaling of different data was done with a locally developed software. The x-ray patterns were analysed with reasonable agreement factor using the *FullProf* software.

## IV. EXPERIMENTAL RESULTS AND DISCUSSION

### a. Polarization reversal

Figure 1 shows the P(E) relations at several fixed temperatures, for the compositions $x$ = 0.2, 0.3, and 0.4, respectively. The analysis of the P(E) relations allows us to determine the temperature dependence of the remanent polarization, which is displayed in Figure 2.

For the compositions $x$ = 0, 0.05, 0.10 and 0.15 only linear P(E) relations were obtained in the 6 K < T < 50 K temperature range. This result shows that no polar domain reversal exists in these compositions, suggesting that these compounds do not exhibit ferroelectric phases, as it was reported earlier.[23,24] For the other studied compositions, linear P(E) relations are observed for T > $T_{AFM\text{-}2}$.

We start our discussion with the results obtained on the $Eu_{0.8}Y_{0.2}MnO_3$ sample (Figure 1(a)). As the temperature decreases from $T_{AFM\text{-}2}$ = 27 K towards $T_{AFM\text{-}3}$ = 23 K, hysteresis loops can be detected, with an unusual elongated shape. A limited value of the saturation of the electric polarization could not be achieved, even for electric fields of up to 15 kV/cm, due to its easily polarizable character. On further temperature decreasing, linear P(E) relationships are retrieved. This result reveals that the ferroelectricity in $Eu_{0.8}Y_{0.2}MnO_3$ is intrinsic in the AFM-2 magnetic phase, and not in the AFM-3 phase, as previously suggested.[21,33]

Figure 1(b) shows the P(E) relations obtained in $Eu_{0.7}Y_{0.3}MnO_3$. As it can be seen, hysteresis loops appear just below $T_{AFM\text{-}2}$ = 26 K. However, the most remarkable result is the retrieval of linear P(E) relations below 18 K suggesting a further phase transition, which to the best of our knowledge has not yet been reported for this composition below $T_{AFM\text{-}2}$. A carefully analysis of the dielectric loss as a function of temperature reveals a tinny anomaly at 18 K, reinforcing the possibility of a third and new phase transition in $Eu_{0.7}Y_{0.3}MnO_3$.[33]

For the $Eu_{0.6}Y_{0.4}MnO_3$ compound, hysteresis loops are observed between $T_{AFM\text{-}2}$ = 24 K and 6 K (Figure 1(c)), but the temperature behaviour of the remanent polarization is not monotonous (Figure 2). As the temperature decreases from $T_{AFM\text{-}2}$, the remanent polarization increases, reaching a local maximum at around 21 K, corresponding to the rise of the electric polarization along the *c*-axis.[23,24] On further cooling, $P_r(T)$ decreases, attaining a minimum at 19 K, and then, an increase of $P_r(T)$ is observed down to 15 K. The increase of electric polarization just below 19 K is associated with the appearance of the component of the electric polarization along the *a*- axis, which is larger than the *c*-component.[23,24] Cooling the sample from 15 K down to 6 K, a decrease of the $P_r(T)$ is observed. These results show that $Eu_{0.6}Y_{0.4}MnO_3$ is ferroelectric below $T_{AFM-2}$, as previously reported, but the temperature dependence of the remanent polarization has no counterpart in the temperature behaviour of the electric polarization reported in current literature.[23,24] Among the studied compositions, the $Eu_{0.6}Y_{0.4}MnO_3$ exhibits the largest maximum value of the remanent polarization, which increases from 13.3 $\mu C/m^2$, for x = 0.2, to 55.1 $\mu C/m^2$ for x = 0.4.

## b. Structural refinement

The powder x-ray diffraction patterns obtained for the $Eu_{1-x}Y_xMnO_3$ ceramics, with $x$ = 0, 0.2, 0.3, and 0.4, investigated in the temperature range 5 – 300 K, were refined using Rietveld analysis in order to follow the evolution of relevant structural parameters. All patterns revealed the orthorhombic structure P*bnm* above $T_N$. A typical result of this analysis is shown in Figure 3.

Below or above $T_{AFM-2}$ the x-ray patterns are qualitatively similar. As an example, Figure 4 exhibits the x-ray patterns recorded in $Eu_{0.8}Y_{0.2}MnO_3$ at 15 K, 23 K, and 35 K, respectively. No additional reflections could be detected in the ferroelectric phase. Taking into account the value of the remanent polarization presented to above, the average ion displacement is estimated to be about $10^{-5}$ Å, which is typical for improper ferroelectric phases, as it is the case in rare-earth manganites. No satellites associated with the modulated structure have been detected, as well as peak splitting associated with symmetry reduction. In these conditions, we have no reasons to use a space group different from P*bnm*, that, for the polar state, has to be considered as a high-symmetry approximant of a non-centrosymmetric structure. The corresponding structural distortions of P*bnm* are, in fact, too small to be detected by powder diffraction method Nevertheless, this space group allows us to follow the average structural features associated with the magneto-elastic coupling across the magnetic phase transitions, as it will be further described. Table I presents the reliability factors obtained from the final refinement of the atomic positions, for the compositions x = 0, 0.2 and 0.3, at two different temperatures.

## c. Lattice parameters

### c.1. Concentration dependence

The concentration dependence of the lattice parameters *a*, *b*, and *c*, as well as of the unit cell volume *V* is depicted in Figures 5(a) and 5(b), for T = 250 K and T = 10 K, respectively. Lines in both figures are guides for the eye. For all samples, the lattice parameters fulfil the $c/\sqrt{2} < a < b$ relation, which is characteristic of the so-called O' structure, typically found in other rare-earth manganites presenting distortions in the octahedral environment of the $Mn^{3+}$ ions, associated with both a strong Jahn-Teller distortion of the $MnO_6$ units, and orbital ordering.[39] The increase of Y concentration leads to a decrease of lattice parameters, as expected from the decrease of the effective A-site size. At 250 K, a linear relation of both lattice parameters and volume with Y-concentration is observed. The anisotropic character of the $Eu_{1-x}Y_xMnO_3$ system is clear from Figure 5(a). The slope of the straight line *b*(*x*) parameter ($\Delta b/\Delta x$ = -0.029±0.001Å) is smaller than the ones of *a*(*x*) and *c*(*x*) ($\Delta a/\Delta$x = -0.08876±0.001 Å, $\Delta c/\Delta$x = -0.108±0.001 Å), as it has been observed in other rare-earth manganites. This kind of behaviour has been attributed to the tilting of $MnO_6$ octahedra in P*bnm* perovskite structure, for which the distortion driven by a reduction in the A-site volume leaves *b* slightly *x* dependent. Contrarily to the linear behaviour observed at room temperature, at 10 K, a non-linear relation between *x* and both lattice parameters and volume is found (see Figure 5(b)).

### b.2. Temperature dependence.

The temperature dependence of *a*, *b*, and *c* is shown in Figures 6(a)-(d), for *x* = 0, 0.2, 0.3 and 0.4, respectively. As expected, all parameters decrease monotonously as the temperature decreases towards $T_N$. The temperature behaviour of each lattice parameter above $T_N$ was analysed in the framework of the Debye thermal expansion equation, according to:[40]

$$\ell(T) = \ell_o + L_a U\left(\frac{\theta_D}{T}\right), \qquad (1)$$

where the Debye function is defined as:

$$U\left(\frac{\theta_D}{T}\right) = 9RT\left(\frac{T}{\theta_D}\right)^3 \int_0^{\theta_D/T} \frac{x^3}{e^x-1} dx. \qquad (2)$$

Here, $\ell$ is the lattice parameter, $\ell_o$ the lattice parameter at T = 0 K, $U$ stays for the average thermal energy, $L_a$ is a proportionality constant, $\theta_D$ denotes the Debye temperature, and $R$ is the gas constant. Within the approximation used in this model, the thermal expansion coefficient is proportional to the specific heat $C_v$, and, consequently, the lattice parameter turns out to be directly proportional to the average thermal energy $U$, just as in Eq. (1). The smooth curves in Figure 6 are the best fits of the equation (1) to the experimental data, above the Néel temperature. All data recorded above $T_N$ are well described by equation (1). Although the lattice parameters are all of the same order of magnitude, their mean thermal expansion coefficients ($<\alpha>$) depend on both direction and concentration. The coefficient $<\alpha>$ was calculated above $T_N$, and the corresponding values are listed in Table II. The anisotropic character of the samples is again reflected in the differences of the mean values of the thermal expansion. Among the three crystallographic directions, the mean value of the thermal expansion is systematically smaller along the *b*-direction than the other two, which exhibit quite similar values.

Significant deviations from the high-temperature Debye behaviour are observed below $T_N$, which are correlated with the onset of the incommensurate magnetic ordering. The temperature dependence of the lattice parameters is strongly dependent on $x$. The magnetic phase sequence, taking place at lower temperatures, is revealed by small, but distinct anomalies in the temperature dependence of the lattice parameters (see inserts of Figures 6(a)-(d)). As the Y-content increases from $x = 0.2$ to $x = 0.4$, the anomalies of the lattice parameter become larger, attaining for x = 0.4 variations of about 0.002 Å between $T_N$ and $T_{AFM-2}$ (see Figure 6(d)). For this composition, minimum values of the lattice parameters occur at around $T_{AFM-2}$, clearly marking the transition from the incommensurate antiferromagnetic order to the *bc*-cycloidal magnetic structure.

The anomalies observed in the temperature dependence of the lattice parameters across the magnetic phase transitions, confirm the existence of a coupling between spins and lattice, in agreement with the results recently reported in a Raman work on this system.[21] The anomalies observed in the temperature dependence of the lattice parameters are a direct consequence of the atomic displacements, taking place at the onset of a new magnetic arrangement, as will be seen in the following section.

### c. $MnO_6$ deformations and spin-lattice coupling.

The temperature dependence of Mn-O1, Mn-O21, Mn-O22 bond lengths, and the Mn-O1-Mn bond angle is depicted in Figures 7, 8 and 9, for the compositions $x$ = 0.2, 0.3, and 0.4 respectively. The temperature dependence of both bond lengths and bond angle shows significant $x$-dependence, and their anomalies evidence the magnetic phase sequence at low temperatures for each investigated $x$-value.

In the 10 K – 80 K temperature range, for each particular value of $x$, the Mn-O21 and Mn-O22 bond lengths exhibit the largest temperature variations, which are similar for both distances and increase with $x$. The length variation of both Mn-O21 and Mn-O22 bonds is almost the same. For $x$ = 0.2, the maximum Mn-O2 length variation is about 0.02 Å, while, for $x$ = 0.4, this value is about 0.08 Å. The change of the Mn-O1 bond length ranges from 0.01 Å (for $x$ = 0.2) to 0.04 Å (for $x$ = 0.4). In a general way, the temperature dependence of the Mn-O1, Mn-O21 and Mn-O22 bond lengths clearly shows anomalous behaviour both at $T_N$ and $T_{AFM-2}$, revealed either by changes of slope or sudden steps. The variations observed at $T_{AFM-3}$ are consistent with the transition revealed by other experimental techniques at this temperature.[23,24] A detailed examination of the Mn-O21 and Mn-O22 bond lengths as a function of the temperature shows that, except for $x$ = 0.2 in the AFM-2 phase, these two bonds behave almost oppositely; i.e., when one bond length increases, the other one decreases. Strikingly, the difference between the Mn-O21 and Mn-O22 bond lengths increases with increasing $x$. The larger difference between the Mn-O21 and Mn-O22 bond lengths occurs for $x$ = 0.4 at $T_{AFM-2}$.

The temperature dependence of the tilt angle Mn-O1-Mn also exhibits strong $x$ dependence. For $x$ = 0.2, the Mn-O1-Mn bond angle as a function of the temperature decreases as the temperature decreases towards $T_{AFM-2}$, and it shows a clear change of slope at $T_N$. Below $T_{AFM-2}$, the Mn-O1-Mn bond angle increases and reaches a maximum value at around $T_{AFM-3}$. Unlikely, for $x$ = 0.3, the Mn-O1-Mn bond angle increases as the temperature decreases, showing a maximum at $T_{AFM-2}$, and changing its slope at $T_N$. Finally, for $x$ = 0.4, a broad anomaly is observed at $T_N$. In the incommensurate magnetic phase, the Mn-O1-Mn bond angle decreases, reaching an almost constant value at both $T_{AFM-2}$ and $T_{AFM-3}$. The amplitude of the Mn-O1-Mn bond angle anomaly also increases with $x$, in the 5 K – 80 K temperature range, varying from 1.5º for $x$ = 0.2 up to 4º for $x$ = 0.4, which is a rather large value.

The aforementioned changes of both bond lengths and bond angle of the $MnO_6$ octahedra across the magnetic phase transition temperatures, clearly evidence the existence of a strong coupling between the magnetic order and the lattice.

We should highlight the importance of this result as it definitely confirms assumptions forwarded in previously published works carried out in orthorhombically distorted rare-earth maganites.[8,22-24,33,34] What makes them a very interesting set of materials is the fact that they share a common $GdFeO_3$–distortion, where the tilt angle of the $MnO_6$ octahedra becomes larger when the rare-earth radius decreases. This behaviour is illustrated in Figure 10 for several undoped rare-earth manganites and the $Eu_{1-x}Y_xMnO_3$ doped system. As it can be seen from Figure 10, for undoped rare-earth manganites, as the ionic radius decreases, the Mn-O1-Mn bond angle decreases almost linearly. However, for the $Eu_{1-x}Y_xMnO_3$ system, a significant deviation from the linear behaviour observed for undoped manganites, is detected. It is worth noting that a much steeper slope is observed for the $Eu_{1-x}Y_xMnO_3$ system. Since the slope of the Mn-O1-Mn bond angle as a function of $x$ scales with the degree of competition between both the NN neighbour ferromagnetic and the NNN antiferromagnetic exchanges in the basal *ab*-plane, its phase-diagram has then to exhibit very unique features, which distinguish the $Eu_{1-x}Y_xMnO_3$ system from the others. Such features are apparent out from experimentally mapped phase-diagrams.[22-25,33] The phase diagram of $Eu_{1-x}Y_xMnO_3$ has also been theoretically studied by several authors,[8,9,12, 43,44] who assumed that the $GdFeO_3$-distortion induces and enhances the NNN antiferromagnetic against the NN ferromagnetic exchanges. Though the Hamiltonian model used can fairly reproduce the experimental ($x$,T) phase-diagram,[29,43] it is unable to describe the data shown in Figures 7, 8, and 9, as no term involving the coupling between the spins and the elastic tensor components has been considered.

The aferementioned $MnO_6$ distortions of orthorhombic $Eu_{1-x}Y_xMnO_3$ compounds clearly evidence an interplay between spins and lattice, yielding a large magneto-elastic coupling, which actually manifests itself by anomalies in the temperature behaviour of the lattice vibrations. Thus, we have checked for possible signatures of the lattice deformations across the magnetic phase transitions in the lattice modes. Figure 11 shows the eigenfrequency of the rotational $A_g$ mode of the $MnO_6$ octahedra as a function of the temperature, which scales directly with the Mn-O1-Mn bond angle, being the main order parameter.[21,35] The solid lines represent the best fits of the the purely anharmonic temperature behaviour of the eigenfrequency to the esperimental data for T > 100 K.[21] For $x$ = 0.3 (see Figure 11(a)), a faint anomaly at $T_N$ is apparent, and a negative shift of the eigenfrequency from the high temperature anharmonic behaviour is observed below $T_{AFM-2}$. For $x$ = 0.4 (see Figure 11(b)), well-defined anomalies at $T_N$ and a positive shift from the high temperature anharmonic behaviour are observed at $T_N$ and $T_{AFM-2}$, respectively. The negative (positive) shifts of the eigenfrequency regarding the temperature anharmonic behaviour have been interpreted as depending on the relative strength between the ferromagnetic and antiferromagnetic exchange interactions, associated with the eigenmode being considered. [21] The above referred shifts are in good agreement with the weak ferromagnetic character of the $Eu_{0.7}Y_{0.3}MnO_3$, and with the antiferromagnetic character of $Eu_{0.6}Y_{0.4}MnO_3$. The aforementioned anomalies in the temperature dependence of the eigenfrequency of the lattice mode provide a clear evidence for the significant role played by the spin-phonon coupling mechanism in these materials.

## V. PHASE DIAGRAM OF $Eu_{1-x}Y_xMnO_3$ for $0 \leq x \leq 0.4$.

The above results obtained from both the analysis of the x-ray diffraction data and the study of P(E) relation enable us to redraw the ($x$,T)-phase diagram of $Eu_{1-x}Y_xMnO_3$ for Yttrium concentrations below $x = 0.45$, namely the limits of the ferroelectric phase. Complementary specific heat, dielectric constant and induced magnetization data were also used to confirm the various phase boundaries. These data, recorded from the same samples used in this work, are already published.[21,33-35]

Figure 12 shows the proposed ($x$,T)-phase diagram for the $Eu_{1-x}Y_xMnO_3$ system, with $0 \leq x \leq 0.4$. The phase boundaries were traced by introducing the phase transition temperatures obtained from the data referred to above. Below the well-known sinusoidal incommensurate antiferromagnetic phase (AFM-1), observed for all compounds, a re-entrant ferroelectric and antiferromagnetic phase (AFM-2) is stable for $0.2 \leq x \leq 0.4$. The ferroelectric character of this phase is established by both x-rays and P(E) data analysis. The decrease of the remanent polarization as the temperature decreases observed for the compositions $x = 0.2$ and 0.3, can be associated with changes of both spin and lattice structures. As no ferroelectric behaviour was found for $0 \leq x \leq 0.3$ down to 7 K, and taking into account the magnetization data, we have considered a unique weakly ferromagnetic phase (AFM-3). Our current data do not provide any additional grounds to split this phase any further. The phase boundary between AFM-2 and AFM-3 phases for $x > 0.3$ could not be traced, since our data do not unambiguously indicate whether a transition to a non ferroelectric phase will occur at temperatures below the lowest measured temperature.

VI. CONCLUSIONS

This work provides a detailed experimental study of both electric polarization and crystal structure of the orthorhombic $Eu_{1-x}Y_xMnO_3$, with $x = 0$, 0.2, 0.3, and 0.4, across the magnetic phase transitions occurring at low temperatures. This study yields two main outcomes. The first one is the existence of a marked magneto-elastic coupling revealed by changes observed in both Mn-O bond lengths and Mn-O1-Mn bond angle at the magnetic phase transitions. These variations are likely to change the electronic orbital hybridization and, in this way, the exchange interactions between Mn spins. As a consequence, this system exhibits a rich phase diagram with very different spin structures and polar properties. Signatures of the lattice deformations across the magnetic phase transitions were evidenced by anomalies in the temperature dependence of the lattice mode involving rotations of the $MnO_6$ octahedra. These anomalies confirm the existence of a significant spin-phonon coupling in these materials.

The second outcome is the existence of a reentrant improper ferroelectric phase for $x =$ 0.2 and 0.3. Though a ferroelectric phase is also stable for $x = 0.4$, there is no experimental evidence that it vanishes at finite temperatures. From these results and along with those ones obtained from other experimental techniques, the corresponding ($x$,T) phase diagram was traced, yet yielding significant differences with regard to other results previously reported. In particular, the trace of a unique non-ferroelectric low temperature AFM3 phase is out forward. Taken together, our experimental findings manifest that oxygen displacements might be strongly involved in formation of the polar state; as a consequence one may expect a shift of phase boundaries at ($x$,T) phase diagram as a function of oxygen isotope composition.

## Acknowledgment

This work was supported by Fundação para a Ciência e Tecnologia, through the Project PTDC/CTM/67575/2006, and by Programme Alban, the European Union Programme of High Level Scholarships for Latin America (Scholarship no. E06D100894BR).

**Figure Captions**

Figure 1. P(E) relations at several fixed temperatures, for the compositions $x$ = 0.2 (a), 0.3 (b), and 0.4 (c).

Figure 2. Temperature dependence of the remanent polarization, determined from the P(E) relations.

Figure 3. Observed, calculated and difference x-ray diffraction patterns for the $Eu_{0.8}Y_{0.2}MnO_3$ compound at room temperature.

Figure 4. x-ray patterns of $Eu_{0.8}Y_{0.2}MnO_3$ recorded at 15 K, 23 K, and 35 K.

Figure 5. Lattice parameters and unit cell volume as a function of yttrium concentration for (a) 250 K, (b) 10 K.

Figure 6. Temperature dependence of $a$, $b$, and $c$ lattice parameters and unit cell volume, for (a) $x$ = 0, (b) 0.2, (c) 0.3 and (d) 0.4. The solid lines represent the best fits of Eq. (1) to the experimental data above 100 K. Insets: expanded view of the temperature dependence of the lattice parameters, in the temperature range from 5 K to 60 K.

Figure 7. Temperature dependence of the (a) Mn-O1, (b) Mn-O21, (c) Mn-O22 bond lengths, and (d) Mn-O1-Mn bond angle of $Eu_{0.8}Y_{0.2}MnO_3$. The vertical dashed lines mark the critical temperatures.

Figure 8. Temperature dependence of the (a) Mn-O1, (b) Mn-O21, (c) Mn-O22 bond lengths, and (d) Mn-O1-Mn bond angle of $Eu_{0.7}Y_{0.3}MnO_3$. The vertical dashed lines mark the critical temperatures.

Figure 9. Temperature dependence of the (a) Mn-O1, (b) Mn-O21, (c) Mn-O22 bond lengths, and (d) Mn-O1-Mn bond angle of $Eu_{0.6}Y_{0.4}MnO_3$. The vertical dashed lines mark the critical temperatures.

Figure 10. Mn-O1-Mn bond angle as a function of the ionic radius of the A-site ion, for the $RMnO_3$, with R = Nd, Sm, Eu, Gd, Dy (closed circles), and for $Eu_{0.8}Y_{0.2}MnO_3$ (open squares).

Figure 11. Temperature dependence of the eigenfrequency of the tilt mode of (a) $Eu_{0.7}Y_{0.3}MnO_3$, (b) $Eu_{0.6}Y_{0.4}MnO_3$. The solid lines represent the best fits of the purely anharmonic temperature behaviour of the eigenfrequency to the experimental data for T > 100 K.

Figure 12. Phase diagram of $Eu_{1-x}Y_xMnO_3$, for $0 \leq x \leq 0.5$.

**Table Captions**

Table I. Reliability factors obtained from the final refinement of the atomic positions, for the compositions x = 0, 0.2 and 0.3, at 35 K and 50 K.

Table II. Mean thermal expansion coefficients ($< \alpha >$) along the three crystallographic directions, calculated above $T_N$, for compositions $x$ = 0, 0.2, 0.3 and 0.4, respectively.

TABLE I

| $x$ | 30 K | | | | 50 K | | | |
|---|---|---|---|---|---|---|---|---|
| | $R_p$ | $R_{wp}$ | $R_B$ | $R_f$ | $R_p$ | $R_{wp}$ | $R_B$ | $R_f$ |
| 0 | 3.66 | 4.86 | 3.67 | 2.22 | 3.68 | 4.78 | 3.86 | 2.50 |
| 0.2 | 1.68 | 2.54 | 1.93 | 1.40 | 1.78 | 2.58 | 2.23 | 1.58 |
| 0.3 | 1.81 | 2.60 | 2.89 | 1.93 | 1.62 | 2.25 | 2.35 | 1.60 |

TABLE II

| $x$ | 0 | | | 0.2 | | | 0.3 | | | 0.4 | | |
|---|---|---|---|---|---|---|---|---|---|---|---|---|
| Parameter | a | b | c | a | b | c | a | b | c | a | b | c |
| $<\alpha> \times 10^{-6}$ $K^{-1}$ | 5.2 | 2.6 | 5.8 | 5.3 | 3.8 | 5.7 | 4.5 | 3.4 | 4.4 | 4.1 | 3.8 | 4.2 |

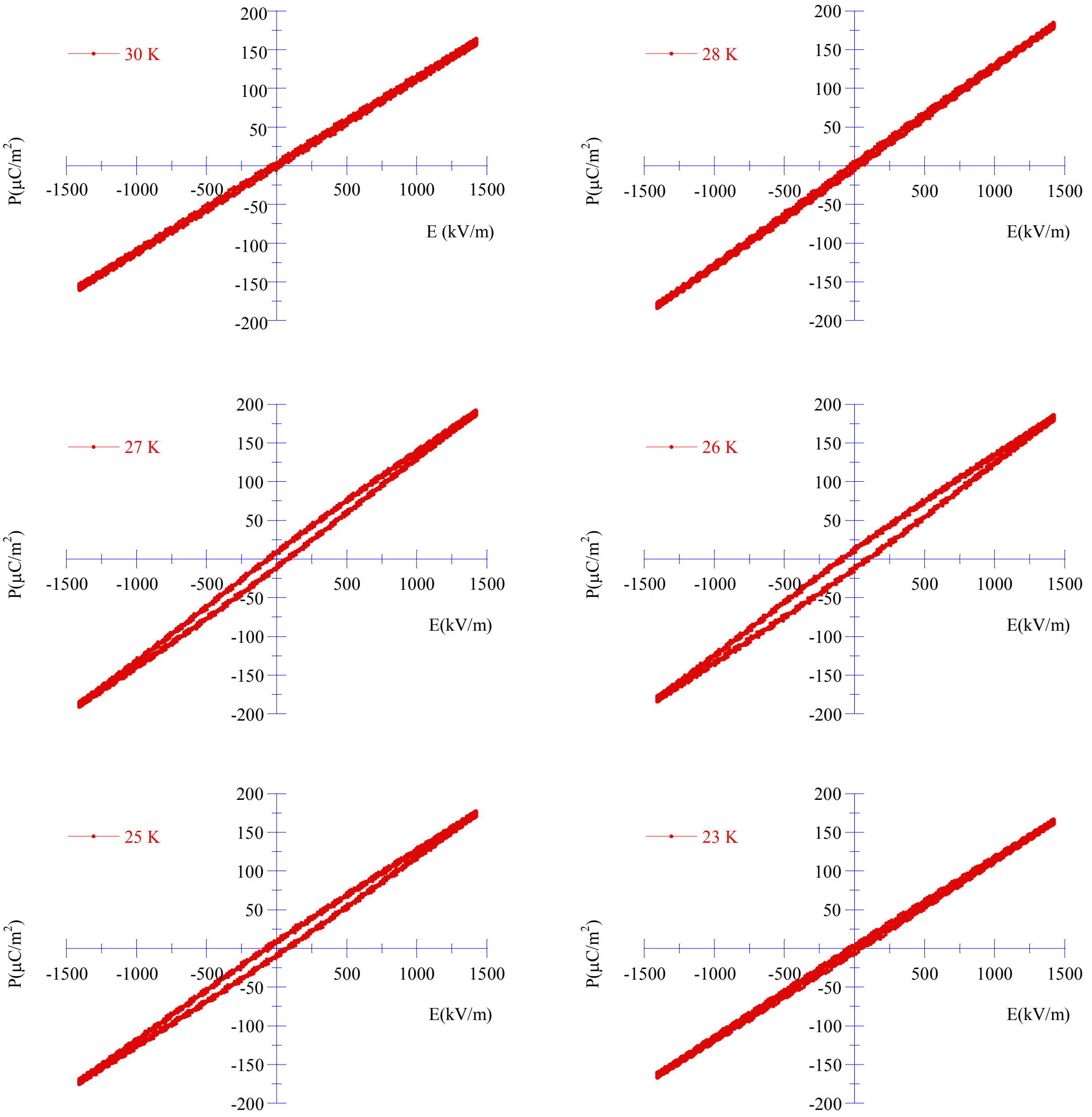

Figure1(a)

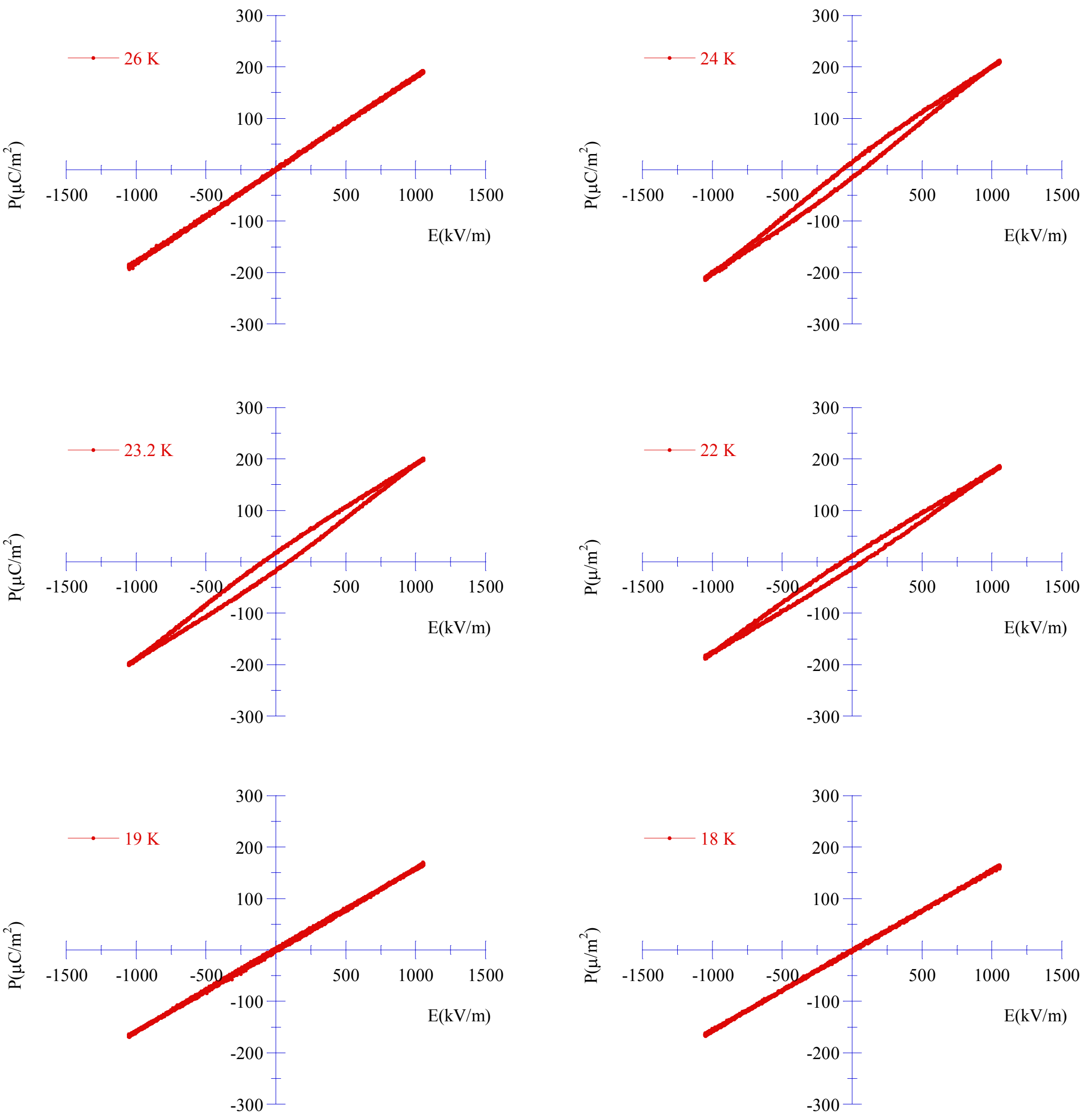

Figure 1(b)

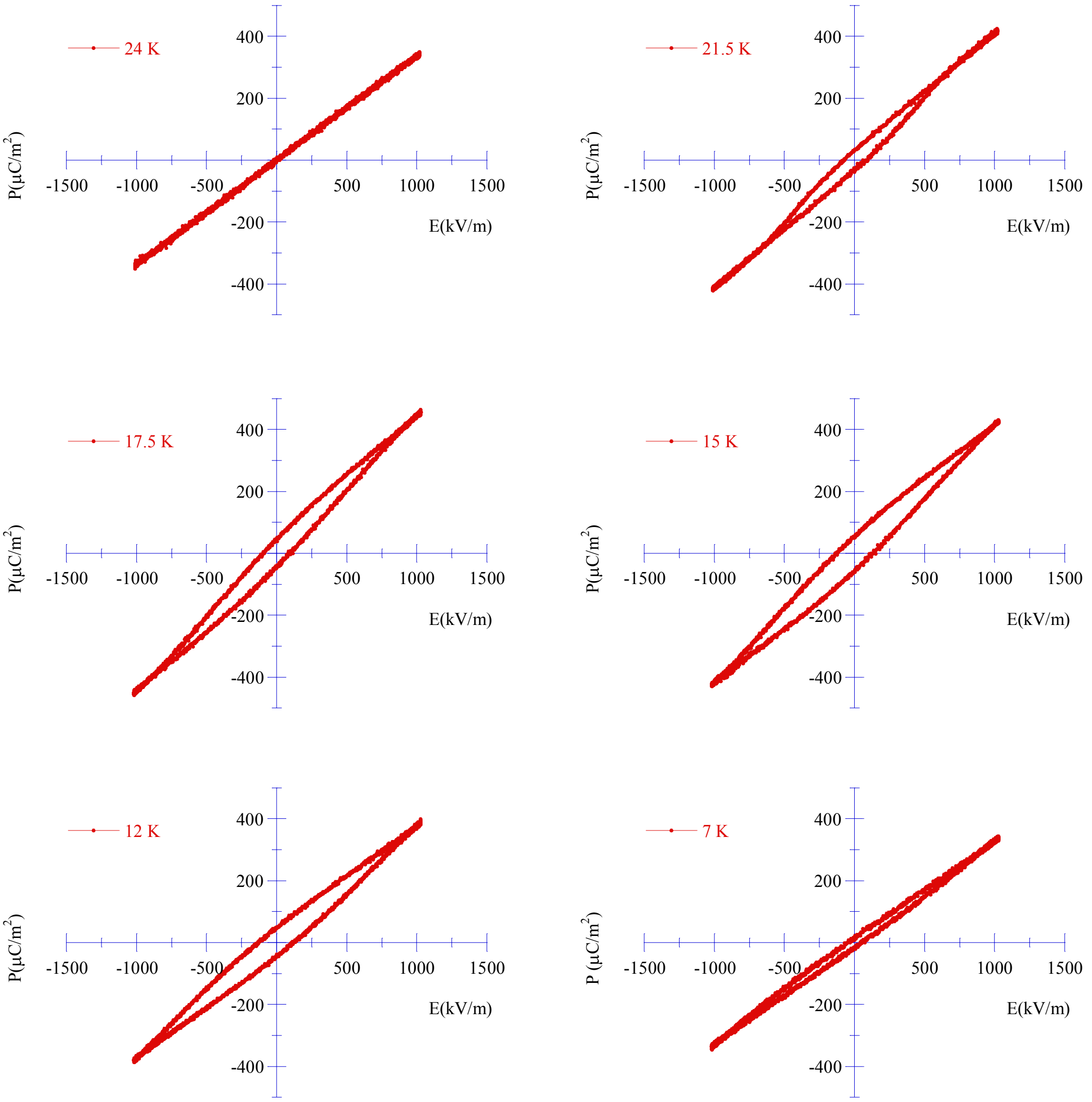

Figure 1(c)

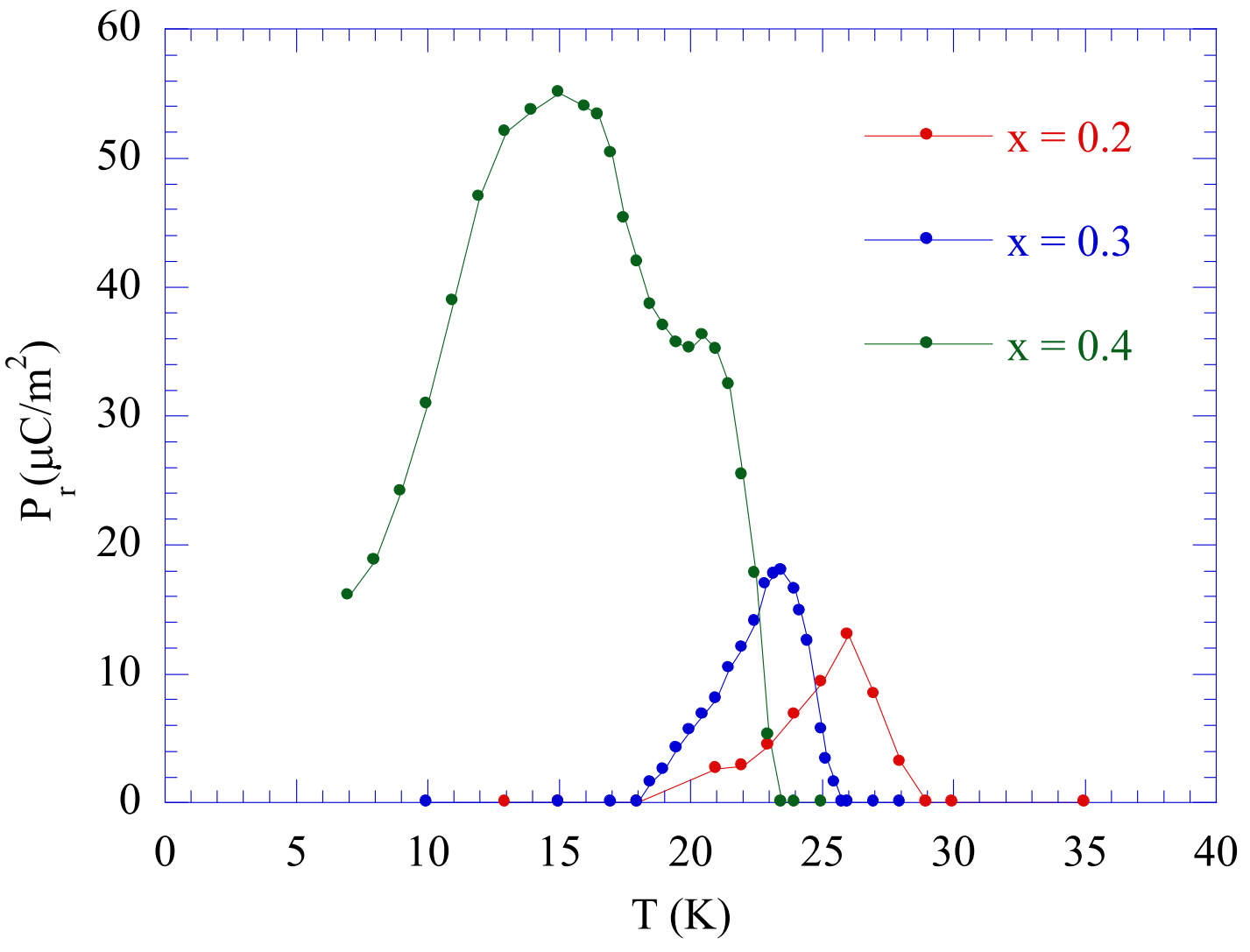

Figure 2

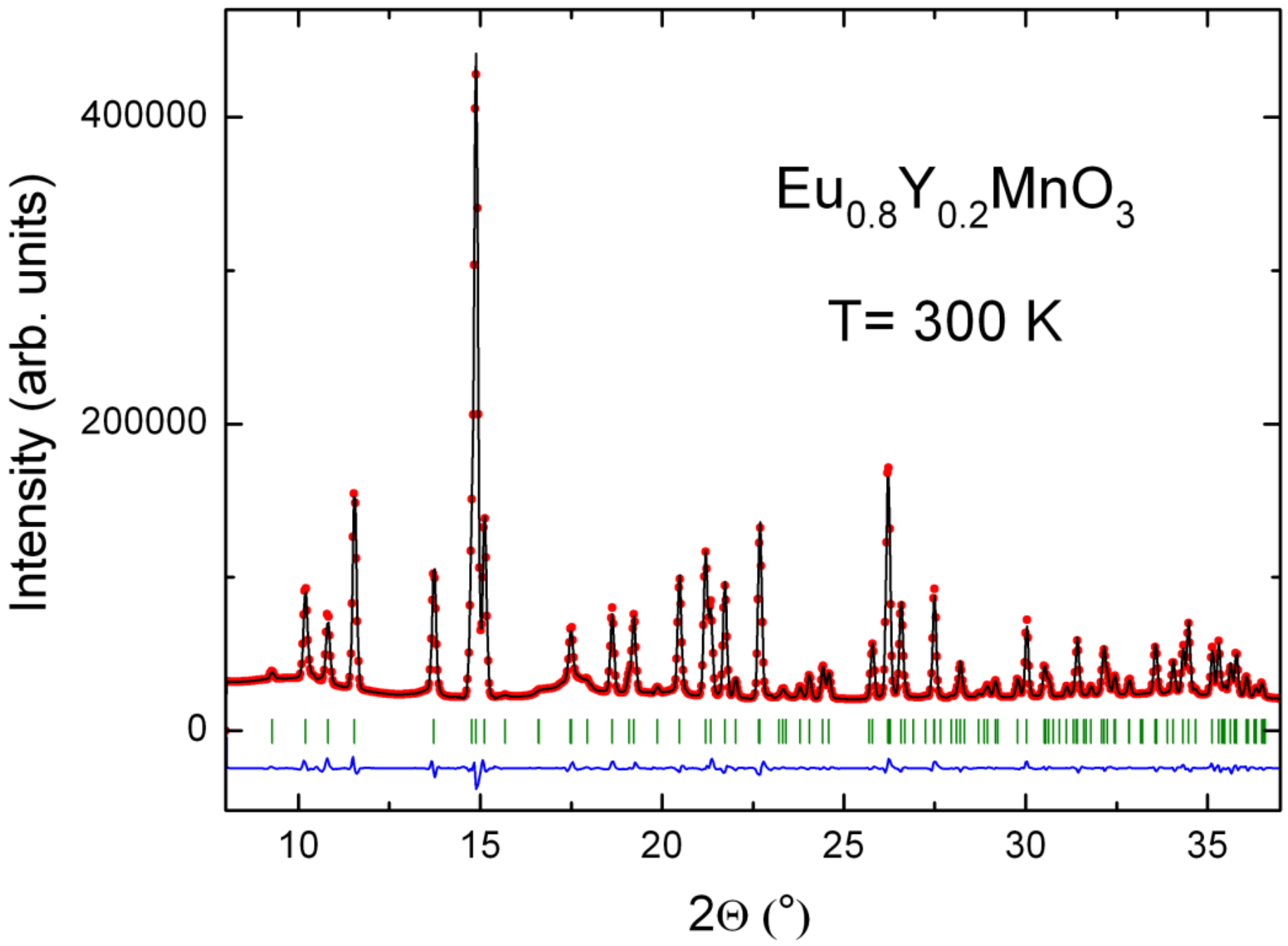

Figure 3

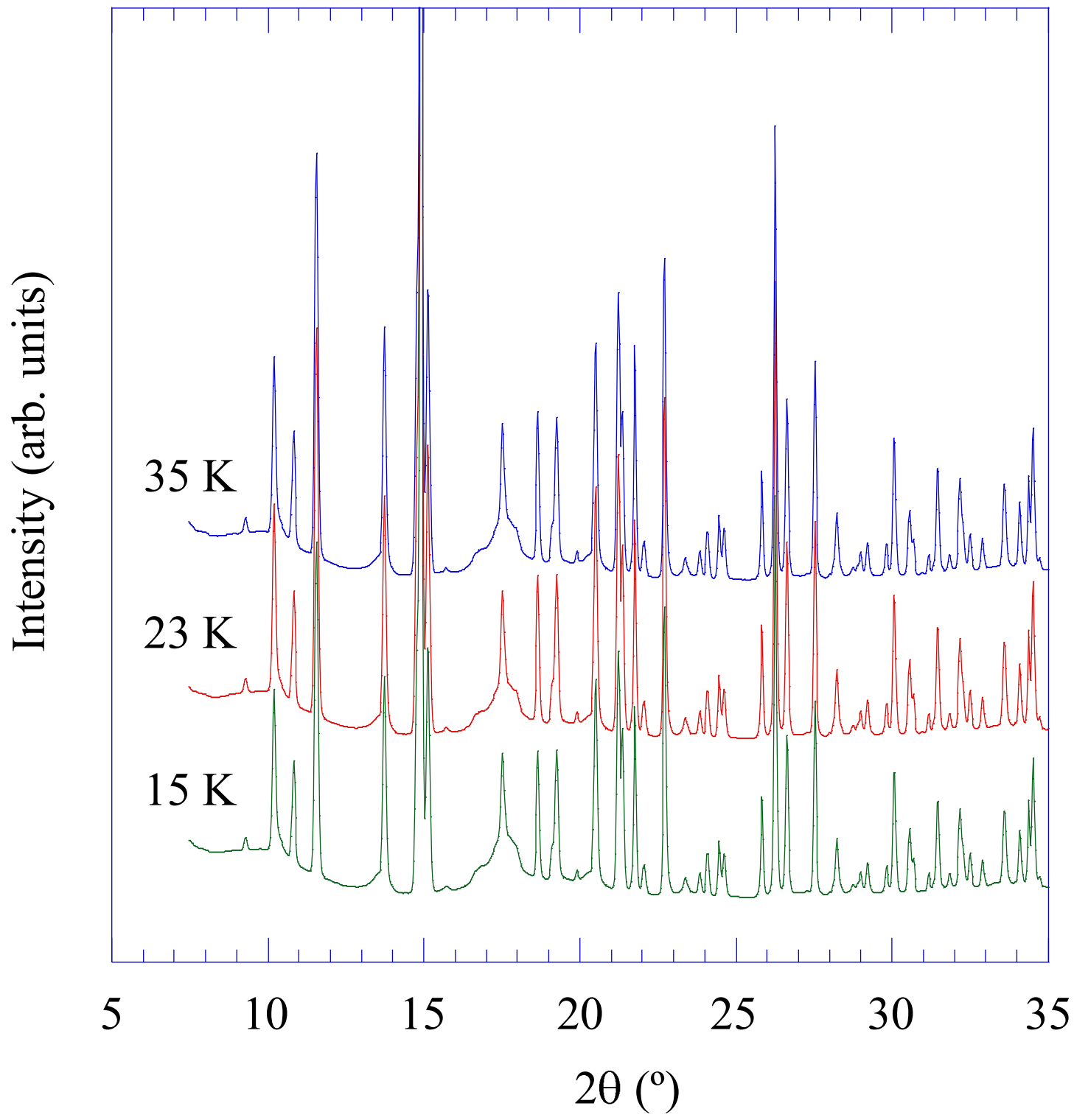

Figure 4

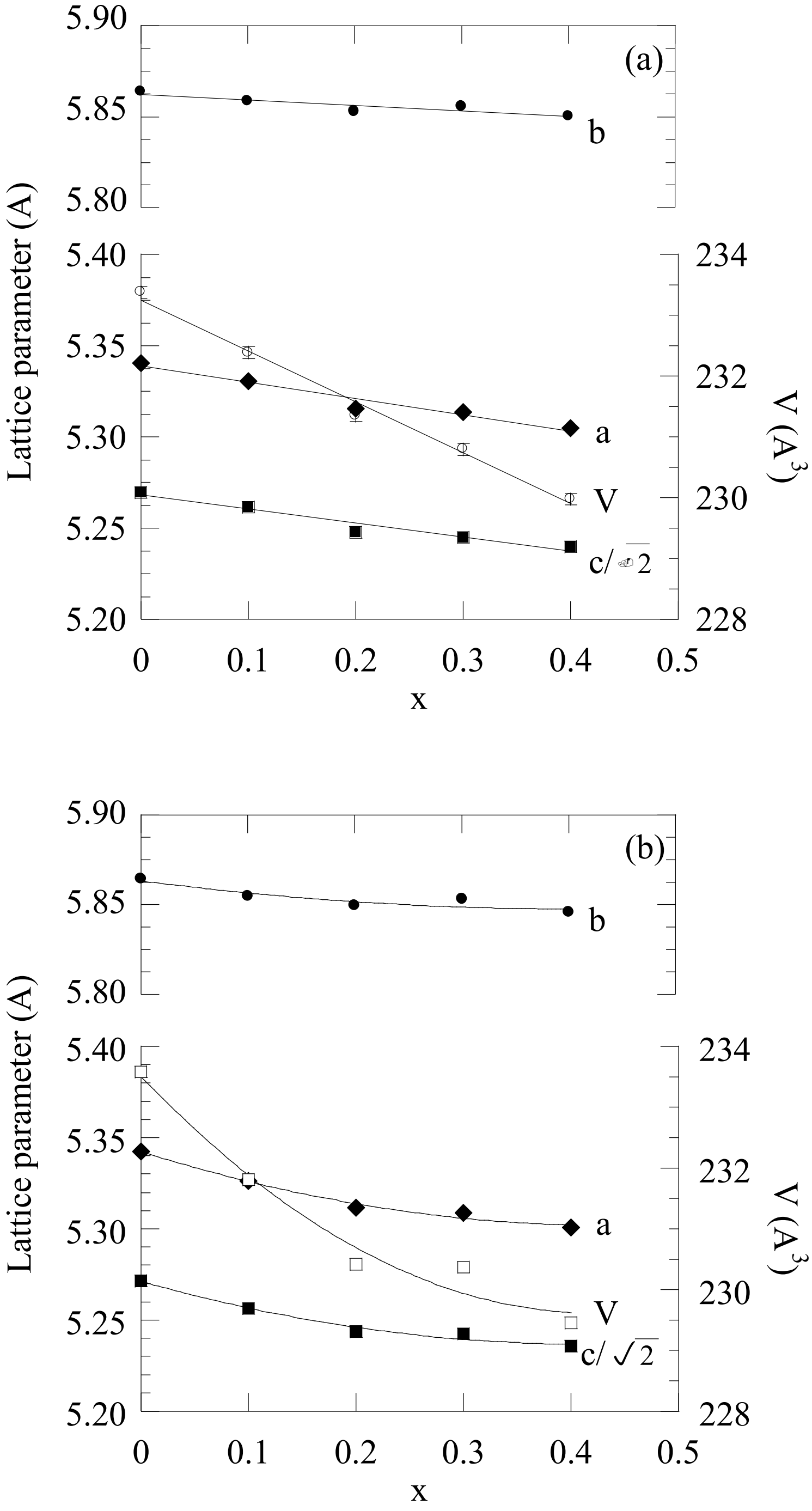

Figure 5

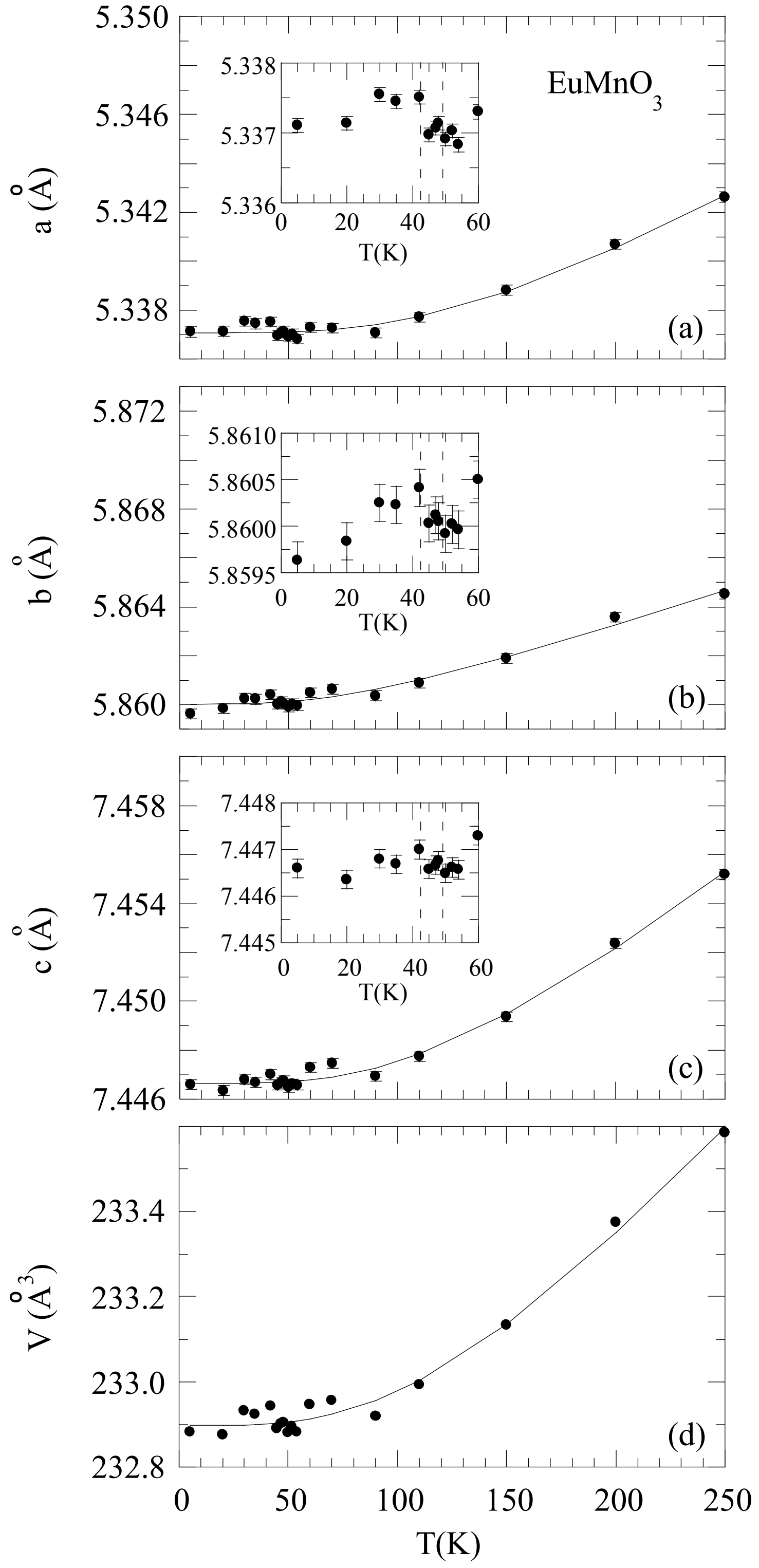

Figure 6(a)

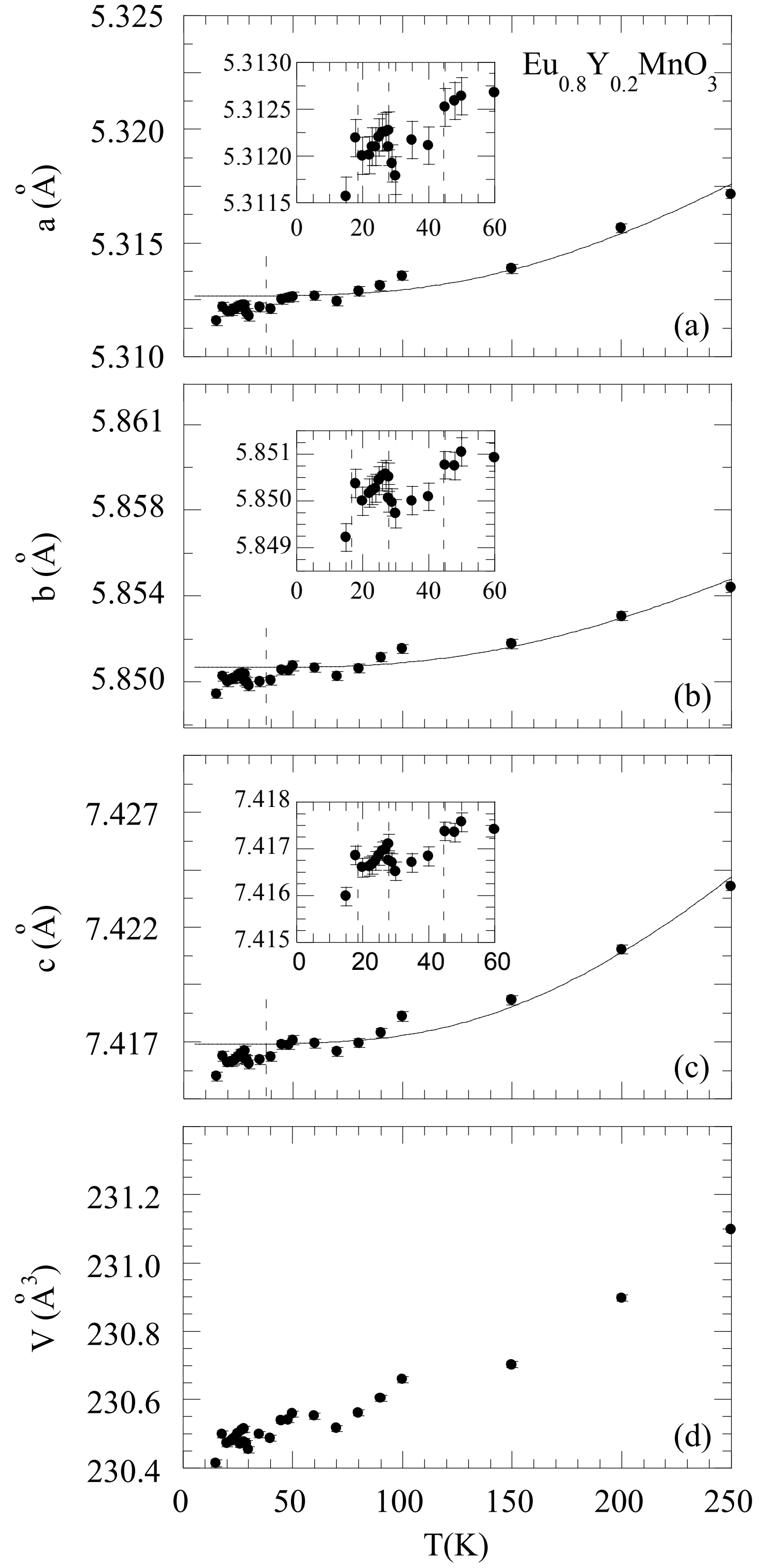

Figure 6(b)

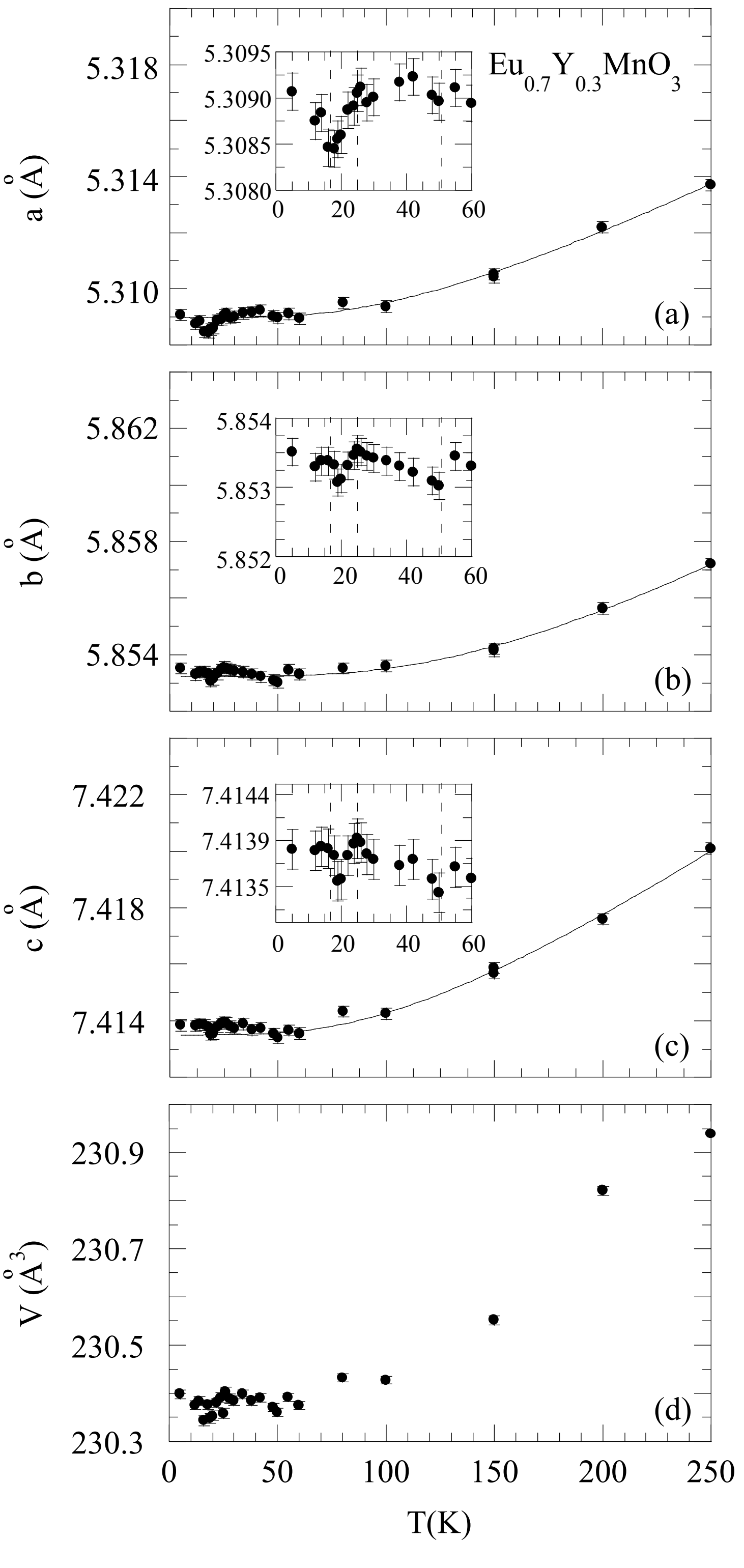

Figure 6(c)

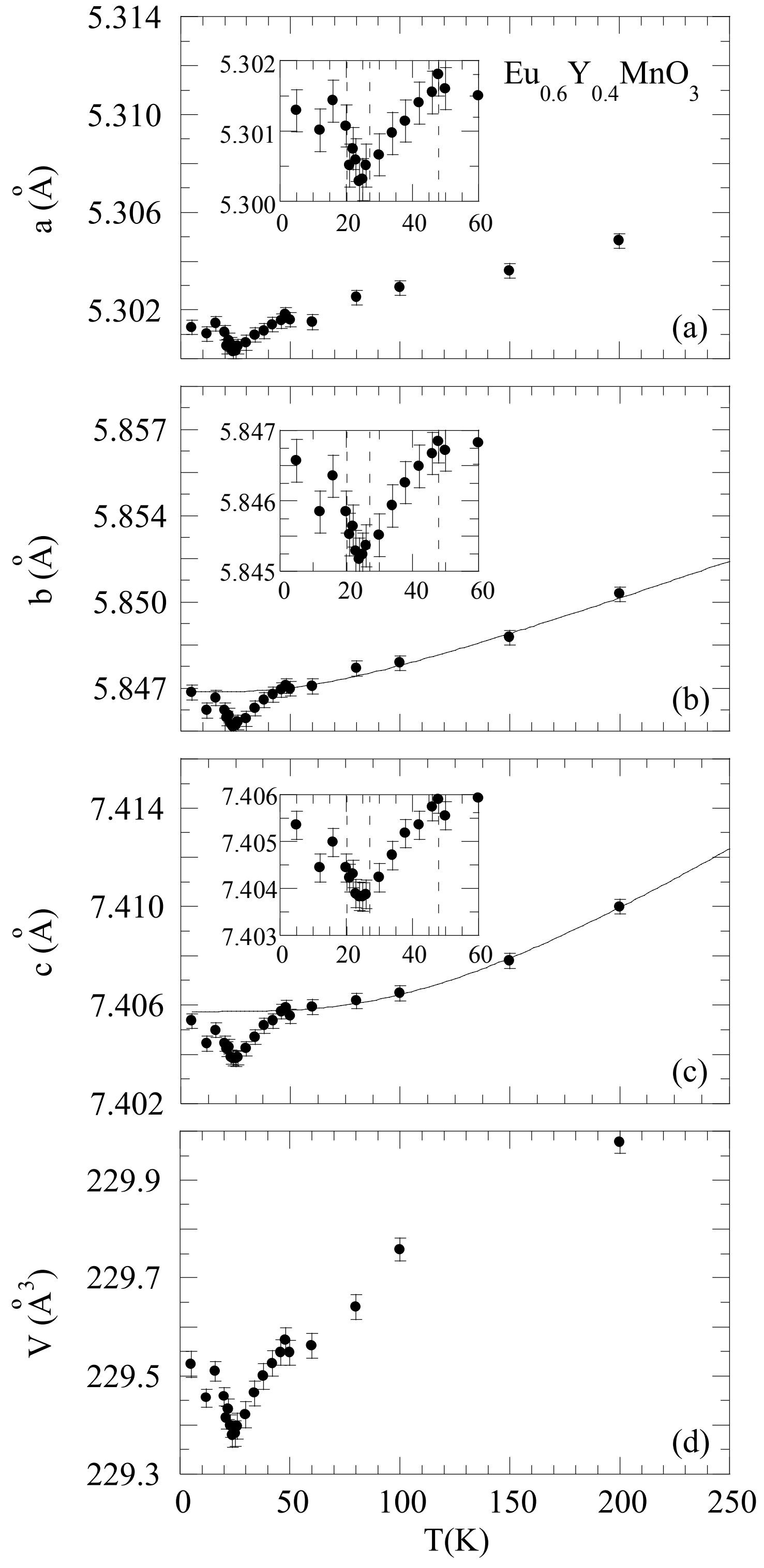

Figure 6(d)

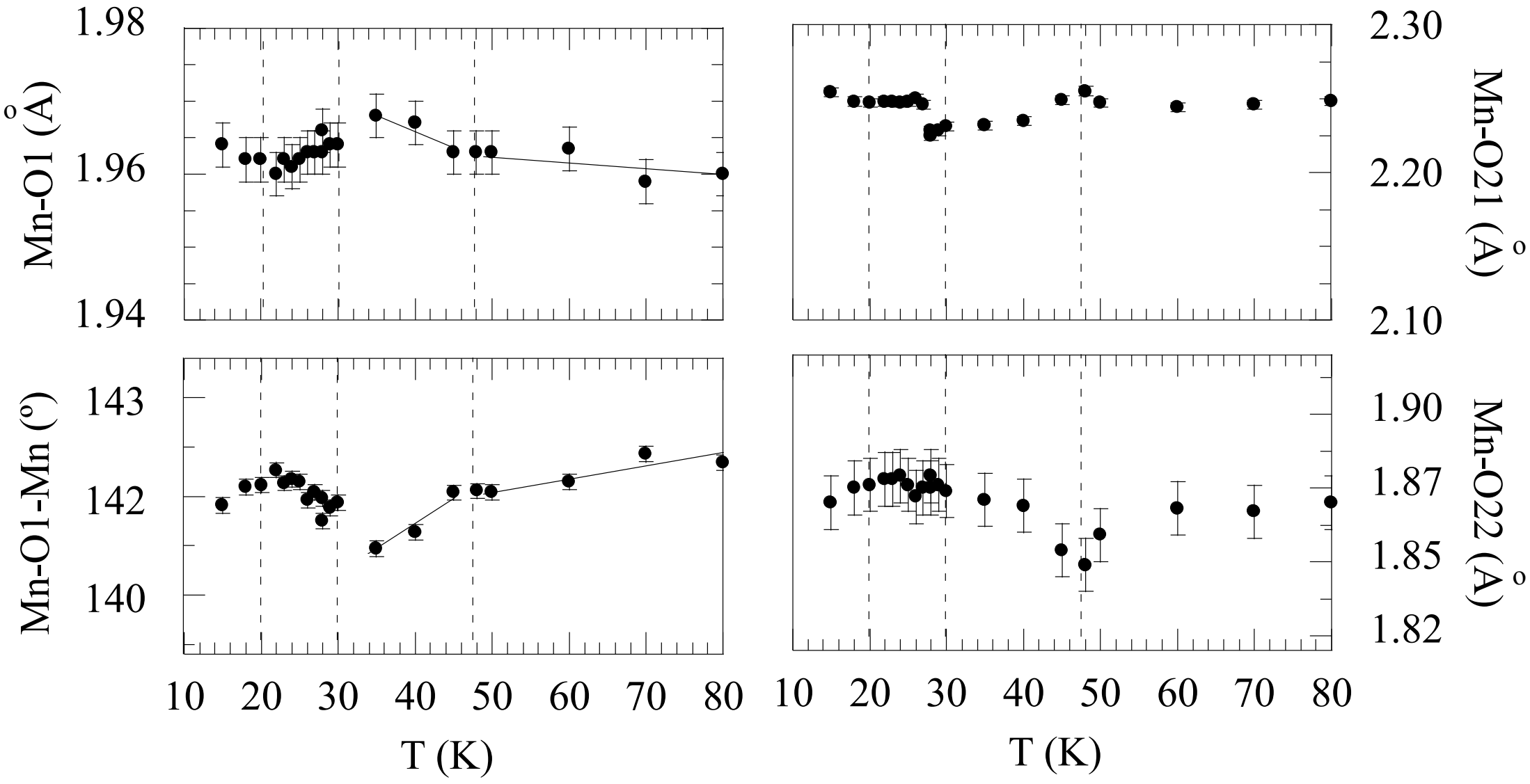

Figure 7

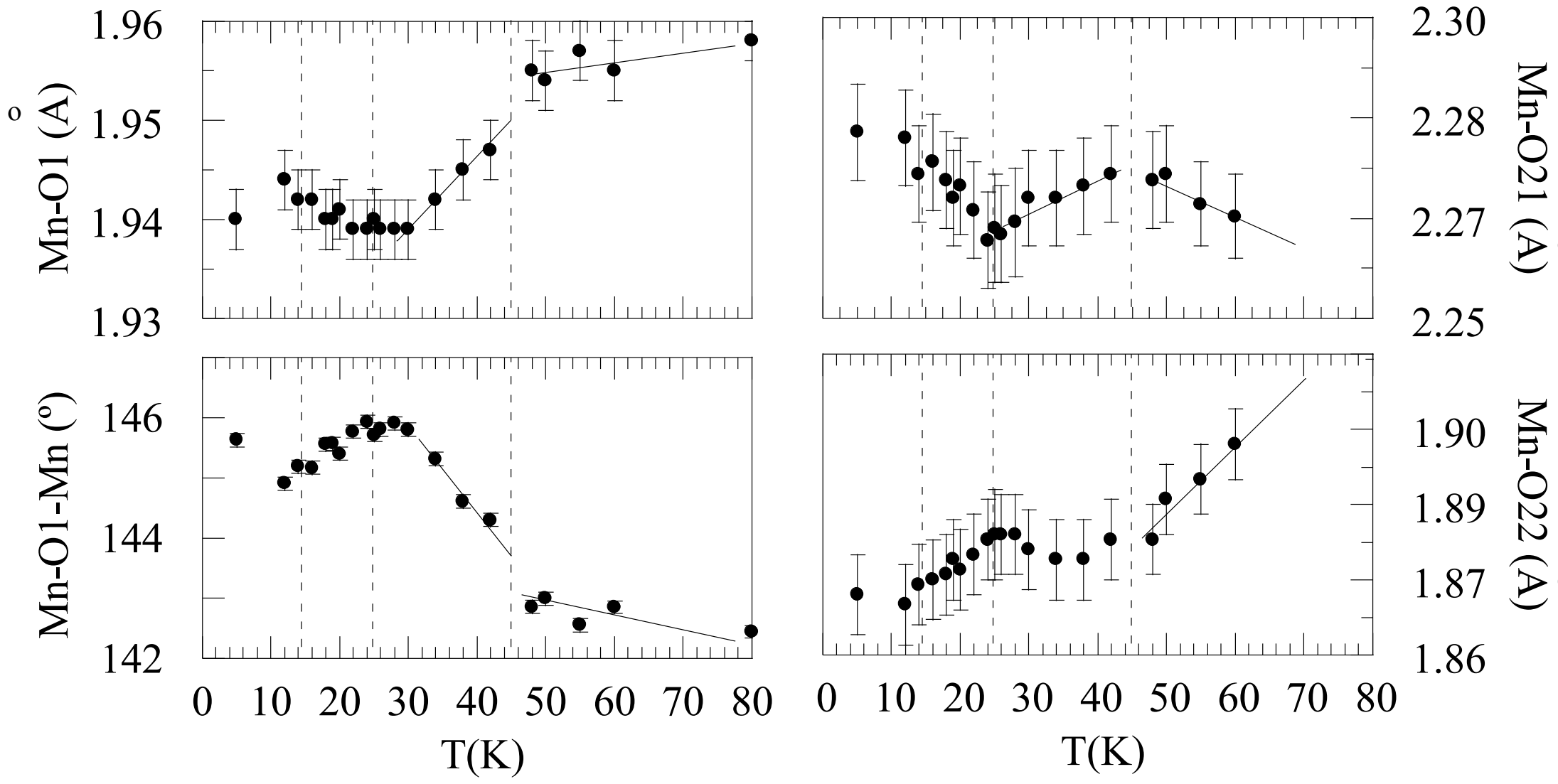

Figure 8

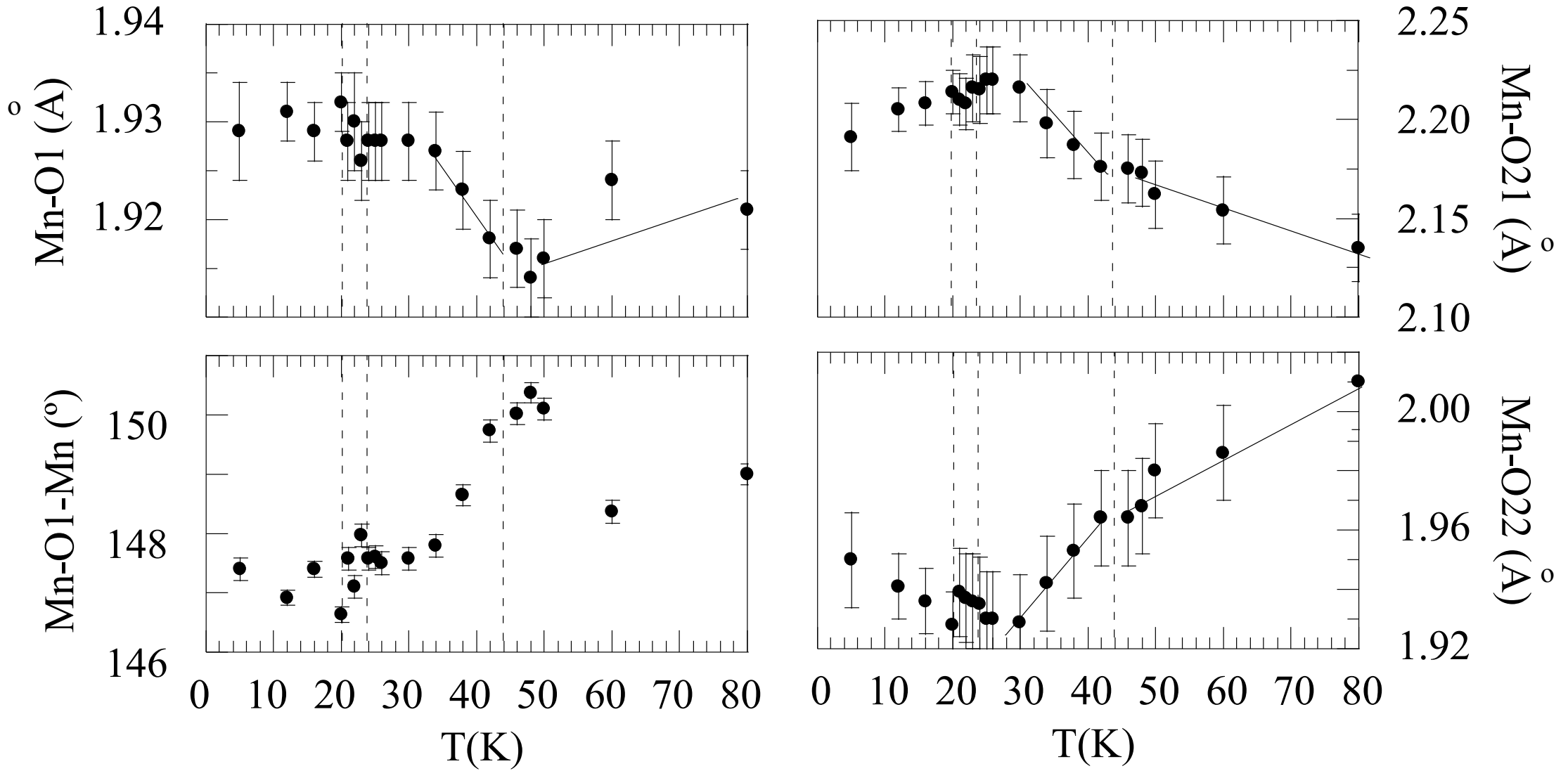

Figure 9

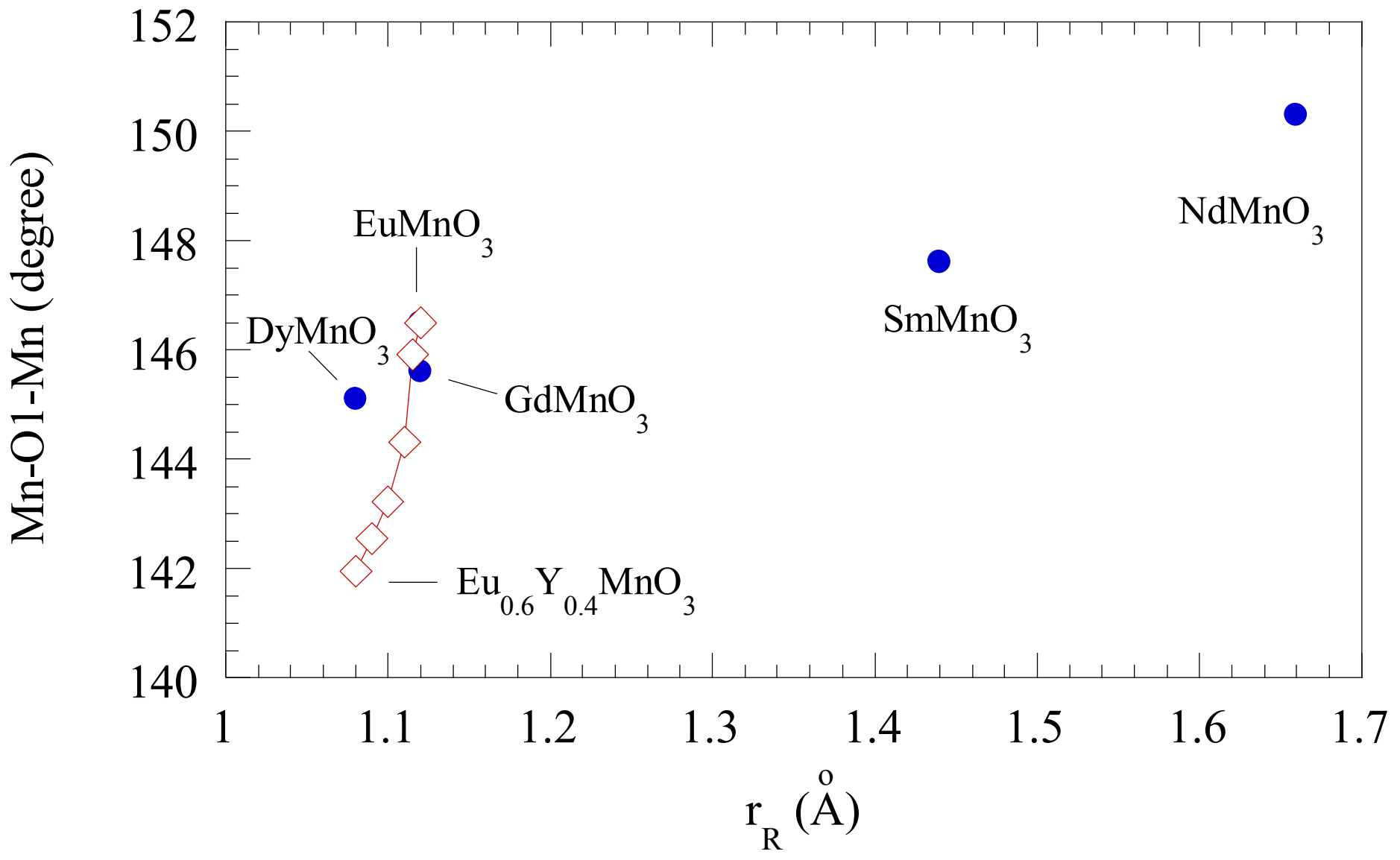

Figure 10

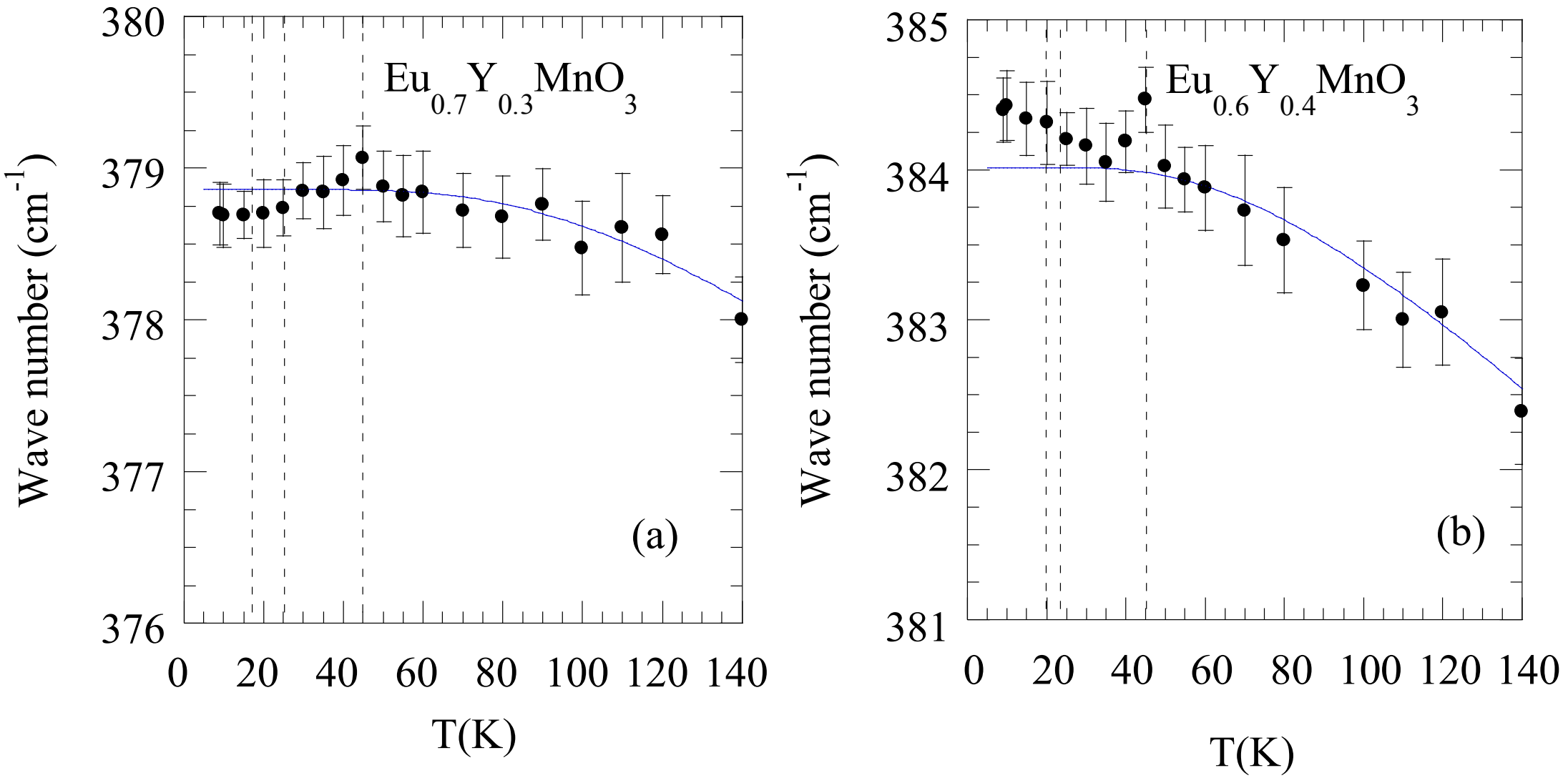

Figure 11

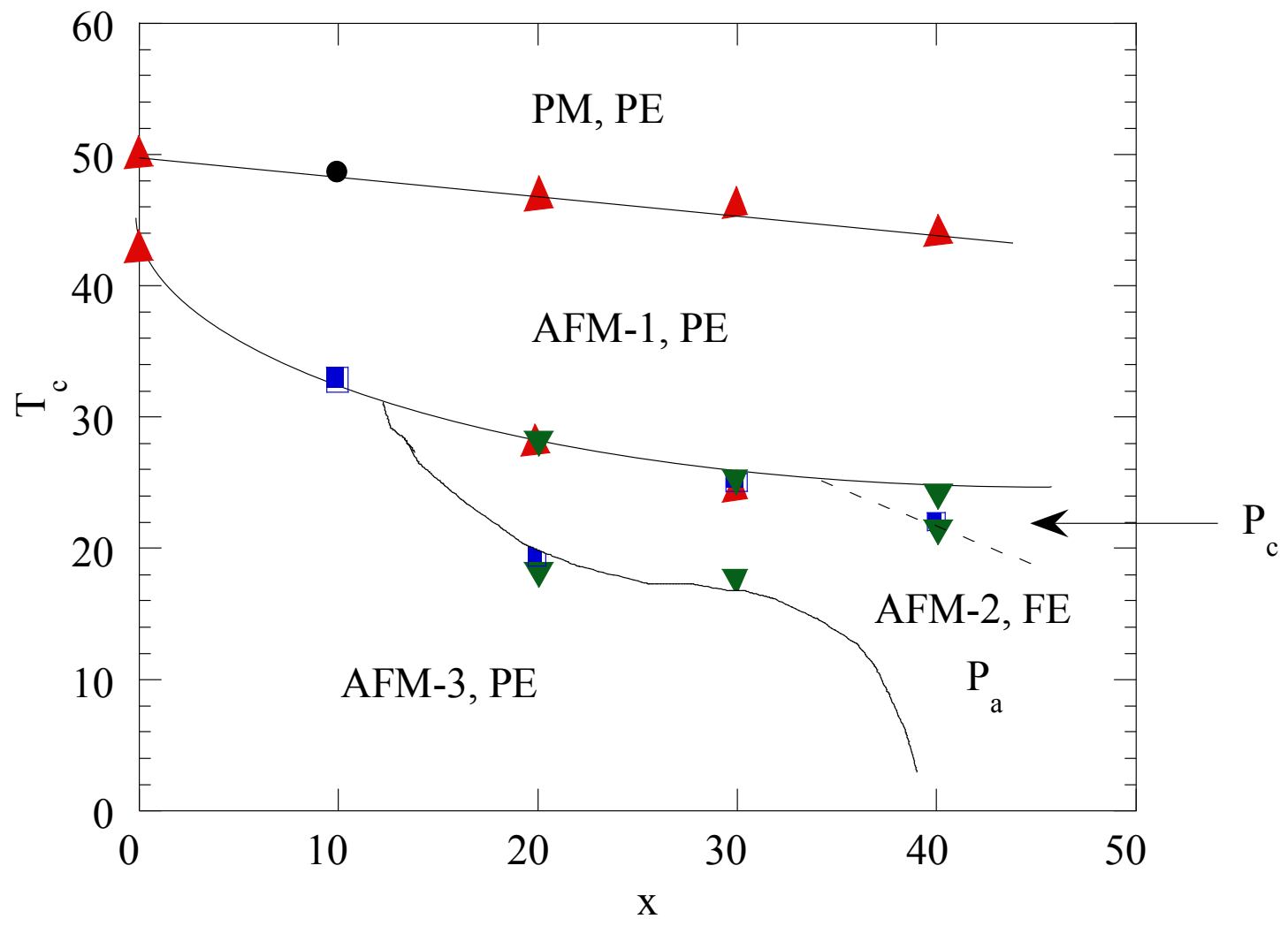

Figure 12